\documentclass[12pt]{amsart}

\newcommand{\R}{\mathbb R}

\newcommand{\x}{\mathbf x}

\newcommand{\CC}{\mathcal C}
\newcommand{\PV}{\mathcal{PV}}

\begin{document}

\title[An attempt to construct dynamical evolution]
{An attempt to construct dynamical evolution in quantum field theory}
\author{A. V. Stoyanovsky}

\thanks{Partially supported by the grant RFFI N~04-01-00640}

\email{stoyan@mccme.ru}

\begin{abstract}
If we develop into perturbation series the evolution operator of the Heisenberg equation
in the infinite dimensional Weyl algebra, say, for the $\varphi^4$ model of field theory,
then the arising integrals almost coincide with the usual Feynman diagram integrals. This
fact leads to some mathematical definitions which, as it seemed to the author,
defined dynamical evolution in quantum field theory in a mathematically rigorous way
using the Weyl algebra.
In fact the constructions of the paper are well defined in perturbation
theory only in one-loop (quasiclassical) approximation.
A variation of the construction is related with the Bogolyubov $S$-matrix $S(g)$.
\end{abstract}
\maketitle

This paper exposes an attempt to construct dynamical evolution in quantum field theory.
The paper is based on the developments
of the previous papers [1--5].
In particular, in [5] a complete mathematical theory of free boson field
is exposed. In the present paper we announce a generalization of results of [5]
to interacting fields.
This paper freely uses the results and notations of [5].

\section{Definition of an attempt of dynamical evolution}

{\bf Definition.} An attempt of dynamical evolution (DE) in quantum field theory on the space-time $\R^{n+1}$ with coordinates
$x^0=t$, $x^1$, $\ldots$, $x^n$ and with boson fields $u^1,\ldots,u^m$ is the following object.
Consider any parameterized space-like surface $\CC$ in $\R^{n+1}$:
\begin{equation}
x^j=x^j(s),\ \ 0\le j\le n;\ \ s=(s^1,\ldots,s^n).
\end{equation}
With this surface one can associate the infinite dimensional Weyl algebra (see its definition
in [5], Subsect. 2.3) of functionals $\Phi(u^i(s),p^i(s))$ on the phase space, where $p^i$ are the
variables conjugate to $u^i$ (see [5]). Denote this Weyl algebra by $W_\CC$. A DE associates
with each pair of parameterized space-like surfaces $\CC_1$, $\CC_2$
an isomorphism of algebras $W_{\CC_1}\to W_{\CC_2}$.
These isomorphisms should constitute an infinitely differentiable family with respect to varying pair $\CC_1$,
$\CC_2$ in the natural sense.
If the surfaces $\CC_1$ and $\CC_2$ present two parameterizations of one and the same space-like surface,
then the isomorphism
$W_{\CC_1}\to W_{\CC_2}$ should coincide with the natural action of a change of variables $s$ on
functionals $\Phi(u^i(s),p^i(s))$. (The variables $u^i(s)$ are transformed like functions, and $p^i(s)$
are transformed like densities.)
The isomorphisms should also satisfy the associativity condition:
for any triple $\CC_1$, $\CC_2$, $\CC_3$ the isomorphism $W_{\CC_1}\to W_{\CC_3}$ should coincide
with the composition of isomorphisms $W_{\CC_1}\to W_{\CC_2}$ and $W_{\CC_2}\to W_{\CC_3}$.
Finally, this family of isomorphisms should be invariant with respect to the given action of the
symmetry group (the Poincare group, or the gauge group) on the space of variables $x,u$.

An {\it attempt of quantization} of a classical field theory given by a variational principle ([5], formula (2))
is a DE depending on the parameter $h$ (which appears in the definition of multiplication in the Weyl
algebra) such that its classical limit as $h\to 0$ coincides with the family of isomorphisms of
Poisson algebras of functionals $\Phi(u^i(s),p^i(s))$ on the phase spaces of space-like surfaces $\CC_1$, $\CC_2$;
these isomorphisms of Poisson algebras are given by the evolution operators of the generalized
canonical Hamilton equations ([5], eq.~(16)).

\section{Perturbation series expansion}

In this Section we will mainly consider so-called attempt of restricted dynamical evolution
(RDE)
involving only flat space-like surfaces $t=const$. By definition, a RDE is a one-parametric
group of automorphisms of the Weyl algebra $W$ of functionals $\Phi(u^i(\x),p^i(\x))$,
$\x=(x_1,\ldots,x_n)$. Any Poincare invariant DE yields a RDE after restriction to flat space-like
surfaces of constant time.

A basic example of DE is the free scalar field, see [5], \S3. The corresponding RDE is given by the
Heisenberg equation (in the notations of loc. cit.)
\begin{equation}
ih\frac{\partial\Phi}{\partial t}=[H_0,\Phi],
\end{equation}
where $H_0$ is the Hamiltonian of a free field. Since $H_0$ is quadratic, we have
\begin{equation}
\frac1{ih}[H_0,\Phi]=\{\Phi,H_0\},
\end{equation}
hence the evolution operator $A_0(t):W\to W$ at the time $t$ is given by the linear symplectic change of variables
given by the evolution operator of the Klein--Gordon equation ([5], eq. (42)). In [5] this evolution operator $A_0(t)$
was symbolically denoted by conjugation by $\exp tH_0/(ih)$ in the Weyl algebra. However, if one attempts
to compute $\exp tH_0/(ih)$, then one immediately realizes that this exponent does not exist in the Weyl algebra,
since already $H_0*H_0$ does not exist. We conjecture that for $t\ne0$, $A_0(t)$ is not an inner automorphism of $W$
(i.~e. not a conjugation by an element of $W$).

Consider now the $\varphi^4$ model as a typical illustration of the perturbation theory method. The classical $\varphi^4$
theory is given by the variational principle ([5], formulas (2),(13)) with the potential term
$V=\frac{m^2}2u^2+\frac g{4!}u^4$, where $g$ is the coupling constant. Consider the Heisenberg equation
in the Weyl algebra
\begin{equation}
ih\frac{\partial\Phi}{\partial t}=[H,\Phi],
\end{equation}
where $H$ is the Hamiltonian of the $\varphi^4$ theory. If we apply usual perturbation expansion to the evolution of this
equation from the time $t_0$ to the time $t_1$, i.~e. if we formally develop into series the symbolical expression
\begin{equation}
\exp\frac{t_1-t_0}{ih}H*\exp\frac{t_0-t_1}{ih}H_0,
\end{equation}
then a computation in the Weyl algebra shows that the perturbation series is given by the usual Feynman diagram integrals
in configuration space, with two differences: 1)~instead of the usual Feynman propagator $1/(p^2-m^2+i\varepsilon)$
we have the propagator $\PV\ 1/(p^2-m^2)$, where $\PV$ denotes the Cauchy principal value; 2)~integration goes
not over the whole space-time $\R^{n+1}$ but over the domain between the planes $t=t_0$ and $t=t_1$.
Hence these integrals are divergent. Let us call them modified Feynman integrals.

Now we make the key assumption which seems reasonable. Assume that one can naturally
renormalize the modified Feynman integrals. Denote by $U(t)$, $t=t_1-t_0$, the renormalized perturbation series,
and by $A(t)$ the corresponding operator in the Weyl algebra:
\begin{equation}
A(t)\Phi=U(t)*(A_0(t)\Phi)*U(t)^{-1}.
\end{equation}
Then the operator $A(t)$ seems a reasonable candidate for the perturbation expansion of the RDE
evolution operator at the time $t$. This operator $A(t)$ seemingly can be symbolically denoted by conjugation
by $\exp tH^{renorm}/(ih)$, where $H^{renorm}$ is the renormalized (infinite) Hamiltonian.

\section{Later remark}

It turned out that the key assumption of the paper about the possibility to renormalize
the expression (5), equal to
\begin{equation}
T\exp\int_{t_0}^{t_1}\!\int gu(t,\x)^4/4!\,dtd\x
\end{equation}
in the Weyl algebra $W_0$ (see [5] for notations), is false. While for the one-loop diagram
with two vertices the renormalization can be performed, for the two-loop diagram with
two vertices one meets problems.
The author is grateful to I.~V.~Tyutin for pointing out this problem.
This is related to the fact that the function
\begin{equation}
g(t,\x)=\left\{\begin{array}{l}
g,\ t_0\le t\le t_1,\\
0,\ \text{otherwise}
\end{array}\right.
\end{equation}
on the space-time $\R^4$ is not smooth. If instead of the integral (7) we take the integral
\begin{equation}
T\exp\int_{-\infty}^{\infty}\!\int g(t,\x)u(t,\x)^4/4!\,dtd\x
\end{equation}
for a smooth function $g(t,\x)$, say, with compact support, then this expression
seemingly can be renormalized in the Weyl algebra (which is similar to the
Bogolyubov--Parasyuk theorem), and the corresponding operator in the Fock space
gives the Bogolyubov $S$-matrix $S(g)$ [8].

Actually,
DE is well defined in perturbation theory in one-loop (quasiclassical) approximation,
i.~e. up to $o(h)$. This is in accordance with results of Maslov and Shvedov [9], who
defined complex germ in quantum field theory using the Bogolyubov $S$-matrix.

All these statements are discussed and proved in the book [10].

A slightly different definition of infinite dimensional Weyl algebra appeared in the paper [6].
For a finite dimensional theory of the Weyl algebra see, for example, [7], \S18.5, and references therein.

\end{document}